\documentclass{article}

\usepackage{fullpage} 
\usepackage{parskip} 
\usepackage{enumitem}

\usepackage{amsmath}
\usepackage{amsfonts}
\usepackage{amssymb}
\usepackage{amsthm}
\usepackage{amstext}

\usepackage{tikz}
\usetikzlibrary{arrows}

\newcommand{\ZZ}{\mathbb{Z}} 
\newcommand{\RR}{\mathbb{R}} 

\newcommand{\OPT}{\ensuremath{\mathop{\mathrm{OPT}}}}

\newcommand{\problem}[1]{\textsc{#1}}  
\newcommand{\is}{\problem{Interval Scheduling}}
\newcommand{\uis}{\problem{Unit-Value Interval Scheduling}}
\newcommand{\ks}{\problem{Knapsack}}
\newcommand{\uks}{\problem{Unit-Value Knapsack}}
\newcommand{\sssp}{\problem{Shortest Paths}}
\newcommand{\nsssp}{\problem{Nonnegative Shortest Paths}}

\newtheoremstyle{definition-bold-title}%
  {}{}
  {}{}
  {\bfseries}{.}
  { }{\thmname{#1}\thmnumber{ #2}\thmnote{ (#3)}}
\theoremstyle{definition-bold-title}

\newtheorem{definition}{Definition}

\usepackage{xifthen}
\newtheoremstyle{specification-style}%
  {}{}
  {}{}
  {\bfseries}{}
  { }{\thmname{#1}\thmnumber{ #2}: \textnormal{\thmnote{ #3}}} 
\theoremstyle{specification-style}
\newtheorem{specification}[definition]{Problem specification}
\newcommand{\problemspec}[6]{
\begin{specification}[#1]
\ifthenelse{\isempty{#2}}{}{\label{#2}} \ 
\begin{description}[style=multiline,leftmargin=2.5cm,align=right,font=\normalfont\textit]
\item[Input:] #3
\ifthenelse{\isempty{#4}}{}{\item[Precondition:] #4}
\item[Output:] #5
\ifthenelse{\isempty{#6}}{}{\item[Side effect:] #6}
\end{description}
\end{specification}
}

\usepackage{algorithm}
\usepackage[noend]{algpseudocode}

\MakeRobust{\Call}
\algdef{SE}[SUBALG]{Do}{EndDo}{}{\algorithmicend\ }%
\algtext*{Do}
\algtext*{EndDo}

\title{From Dynamic Programs to Greedy Algorithms\thanks{This material is based upon work supported by the U.S.\ National Science Foundation under Grant No.\ CCF-2312540.}}
\author{Dieter van Melkebeek}

\begin{document}

\maketitle

\begin{abstract}
We show for several computational problems how classical greedy algorithms for special cases can be derived in a simple way from dynamic programs for the general case: 
interval scheduling (restricted to unit weights), knapsack (restricted to unit values), and shortest paths (restricted to nonnegative edge lengths). Conceptually, we repeatedly expand the Bellman equations underlying the dynamic program and use straightforward monotonicity properties to figure out which terms yield the optimal value under the respective restrictions. The approach offers an alternative for developing these greedy algorithms in undergraduate algorithms courses and/or for arguing their correctness. In the setting of interval scheduling, it elucidates the change in order from earliest start time first for the memoized dynamic program to earliest finish time first for the greedy algorithm.
\end{abstract}

\section{Introduction}

Many an undergraduate algorithms course features \is\ to illustrate the design paradigms of greed and dynamic programming. The unweighted version admits an appealing greedy strategy that can be shown correct via several types of arguments (charging, exchange, greedy stays ahead). The weighted version exhibits a straightforward principle of optimality and demonstrates one of the measures that can be taken to keep the number of subinstances in a dynamic program small -- ordering the input components judiciously. However, the appropriate order for the memoized dynamic program differs from the order in the greedy algorithm. This tends to confuse students and gives some the impression that, even though both algorithms solve the unweighted version, they are not that closely related. Existing expositions convey intuition behind the choice of one or both of the orders, but do not explain where the difference arises and do not highlight the close relationship between the algorithms.

This paper addresses those shortcomings. We derive the greedy algorithm from the dynamic program by simplifying the latter under the preconditions of the former. The change in order follows naturally. We apply a similar strategy to two other common examples, namely \ks, where we derive the greedy algorithm for the unit-value restriction from the dynamic program for the general case, and \sssp, where we derive Dijkstra's algorithm from the Bellman-Ford dynamic program.

\paragraph{The paradigms.} Let us start by reviewing dynamic programming and greed in the context of a discrete optimization problem $\Pi$. We can think of an instance $x$ of size $n$ as consisting of $n$ components, $x_1, \dots, x_n$, and need to make a decision $y_i$ about each component $x_i$ so as to maximize or minimize an objective function $f_x(y_1,\dots,y_n)$ under certain constraints. 

In a \emph{dynamic program}, we pick a particular component, say $x_1$. We consider all possible settings $v_1$ for $y_1$, optimize $f_x$ with the additional constraint $y_1=v_1$, and assign $y_1$ to a value $v^*_1$ that produces the best result. The approach hinges on an efficient reduction from instances with the above additional constraint to proper $\Pi$-instances of size smaller than $n$. Repeated application yields a recursive algorithm for $\Pi$ that can be captured in the Bellman equations, which express the optimal value for input $x$ as the maximum or minimum of terms involving the optimal values for smaller instances. Correctness typically follows in a streamlined fashion. The name of the game is to ensure that the number of subinstances on any given input $x$ remains small, i.e., the number of distinct calls to $\Pi$ that arise throughout the entire recursive process on input $x$. This often requires reformulating the decisions to choose among, and sometimes also reordering the components. If and when the approach succeeds, memoization yields an efficient top-down algorithm for $\Pi$. Alternately, when the subinstances can be described in terms of $x$ and $k$ additional parameters, a $k$-dimensional table with the optimal values of the subinstances can be computed iteratively in a bottom-up fashion, after which a second phase retrieves decisions that achieve the optimal values in a top-down fashion.

A \emph{greedy algorithm} also starts by picking a specific component, say $x_1$, but is able to efficiently figure out an optimal setting $v^*_1$ for $y_1$ without making any recursive calls. It never considers settings for $y_1$ other than $v^*_1$ and proceeds with the remaining components, resulting in a single iterative phase. The order of the components plays a critical role in getting the approach to work. If and when we can find a suitable order, an optimal setting $v^*_i$ for $y_i$ often follows straightforwardly by taking the settings of the prior components into account and myopically assuming that there are no subsequent components. However, coming up with a suitable order and arguing correctness usually require considerable insight and ingenuity.

\paragraph{Approach and instantiations.}
Instead of coming up with a greedy order for $\Pi$ from scratch, we derive it from a dynamic program for a generalization $\Pi'$ of $\Pi$. We repeatedly expand the Bellman equations underlying the dynamic program for $\Pi'$ and exploit the preconditions of $\Pi$ and basic monotonicity properties to simplify the resulting terms and figure out which ones yield the optimal value. We succeed in the following classical settings (see subsequent sections for problem definitions):
\begin{itemize}
\item \is. 
We start from a straightforward dynamic program for the weighted version and derive the classical greedy algorithm for the unweighed version. The weighted version represents a rare example of a computational problem where the order of the components in the dynamic program matters; earliest start time first 
yields the best bound on the number of subinstances. Our derivation of the greedy algorithm for the unweighted version transforms the order in a natural way into earliest finish time first. The monotonicity of the 1-dimensional table of subinstance solution values justifies the transition.
\item \ks. 
We start from the standard dynamic program for the general problem with integral item weights and arbitrary item values. Compared to \is, the subinstances have an additional parameter, namely, the weight limit of the knapsack. The order of the items plays no role in the analysis of the dynamic program. Our derivation for the case with unit values leads to lightest first as the greedy order. The justification follows from the monotonicity of the 2-dimensional table of subinstance solution values with respect to each parameter separately. 
\item \sssp. 
We consider the single-source version. In contrast to the previous two settings, the decision for each component is non-binary, namely, the last edge on a shortest path from the source to the component vertex. We start from the Bellman-Ford dynamic program for digraphs with arbitrary edge lengths. Our approach turns the program into Dijkstra's greedy algorithm under the precondition that the edge lengths are nonnegative. The justification is the same monotonicity property that underpins the traditional correctness proofs of Dijkstra's algorithm, namely, that extending a path cannot make it shorter. 
\end{itemize}

\paragraph{Correctness considerations.}
The alternate derivations of the greedy algorithms implicitly yield correctness arguments that differ from traditional ones and are more modular. The first two examples are ordinarily proven using a charging argument (one-to-one mappings from the greedy selection to any other valid selection) or an exchange argument (massaging any optimal solution into the greedy solution by a finite sequence of local transformations, each of which maintains validity and optimality). 

No such arguments are known for Dijkstra's algorithm but, interestingly, all three examples can be proven correct using a classical strategy that is dubbed ``greedy stays ahead" in \cite{KT}. The critical ingredient in such an argument is a quality measure for partial solutions such that, for full solutions, optimal quality implies optimal objective value. The correctness argument then consists of showing that at any point in time, the quality measure of the partial greedy solution is at least as high as of any other partial solution on the same components. If and when applicable, the quality measure formally captures intuition for why the greedy order leads to an optimal solution. Coming up with such a quality measure is often nontrivial. In each of the examples, an appropriate quality measure follows from the function argument to which the monotonicity property is applied.

\paragraph{Didactic considerations.}
When designing an undergraduate algorithms course, one needs to decide whether to teach dynamic programming before greed (as done in textbooks like \cite{CLRS,E}) or vice versa (as done in textbooks like \cite{KT,DPV,R}). Arguments in favor of the latter include prior familiarity of students with greed as opposed to dynamic programming, and developing simpler algorithms for more restricted settings before algorithms for the general setting. Arguments in favor of dynamic programming before greed include the structured development and wide applicability of dynamic programming versus the ad-hoc character of greedy algorithms and their limited scope, and that correctness proofs for dynamic programs can be streamlined better and tend to be more straightforward. Our alternate approach can be viewed as another argument in favor of covering dynamic programming first, along with the advice to first develop a correct dynamic program for a problem and subsequently see whether the algorithm can be made faster based on greedy considerations that exploit additional structure in the instances of interest.

In my own offerings, I have switched to covering dynamic programming first. I employ the alternate approach, not to derive greedy algorithms from dynamic programs but rather as concrete instantiations of the above advice and to provide alternate correctness arguments in addition to the traditional ones. In the setting of \is, the explanation of the change in order from earliest start time first to earliest finish time first seems to enlighten students. The alternate view can be helpful even when covering greed first, namely, to reconnect to the greedy algorithms for the special cases after presenting the dynamic programs for the general case. 

\paragraph{Related work and open question.}
The paradigms of dynamic programming and greed are studied in several areas, including algorithms and complexity, operations research and management science, and programming languages. The connections reported in this paper ought to be known, but I have been unable to find them in the literature (including textbooks and lecture notes).
The coverage in \cite{CLRS} gets closest in that it also relates dynamic programming and greed for \is\ and for \ks, but the perspective and connections are quite different from the approach here. For \is, \cite{CLRS} studies the unweighted version only, and only considers earliest finish time first, so there is no explanation for the switch in order. Their dynamic program tries every possible interval to reduce an instance into at most two smaller instances; they then argue that picking the first interval is always best. For \ks, \cite{CLRS} simplifies the problem by allowing fractional items rather than restricting to unit values, and explains why the fractional version allows a greedy approach whereas the original problem does not seem to. The presentation derives the greedy algorithm for the fractional variant from the dynamic program for the proper problem. 

For \sssp, some connections between Bellman-Ford and Dijkstra are well-known, but the derivative approach in this paper seems new. The closest presentations  \cite{Denardo03,Bertsekas17,Sniedovich06} view Dijkstra as a modification of Bellman-Ford where the Bellman equations are applied best first (and each only once) instead of breadth first (and each possibly multiple times). They argue correctness of Dijkstra from scratch rather than deriving it from the correctness of Bellman-Ford. 

We note that there are ways in which Dijkstra's algorithm itself resembles a dynamic program. In the single-pair version, whether to include an edge on the shortest path from $s$ to $t$ is not decided as soon as the edge is considered, but in a second phase whose order is the reverse of the greedy order. Arguably, the single-source version that we consider \emph{is} greedy. In every step, a shortest path to some new target $t$ is completed by the choice of a particular final edge.

Much of the existing literature on the paradigms aims at precisely formalizing them. The programming languages community has developed very general formalisms within the framework of category theory to synthesize programs from problem specifications according to various algorithmic paradigms.
A paper with a very similar title to this work builds dynamic programs and greedy algorithms under correctness conditions that take the form of monotonicity requirements in the categorical calculus of relations \cite{BirdM93}. The conditions for a greedy algorithm are stricter than for a dynamic program, but merely express the very existence of a local criterion that guarantees optimality, and the construction presupposes knowledge of the local criterion. As such, whereas the generality of the formalism is suitable for program synthesis, the results provide little clue to an algorithm designer what greedy approaches might work or how to formulate a given computational problem to enable a greedy approach. Proofs of successful instantiations hinge on classical correctness arguments for known greedy algorithms. See Section~\ref{sec:monotonicity} for more details.

In a simpler framework for optimization problems with a decomposable structure, \cite{Lew06} proposes a ``canonical" way of deriving greedy algorithms from dynamic programs, but without correctness guarantees. The canonical local criterion is the straightforward one induced by the decomposition of the objective function. The framework is easy to apply but is more limited than the categorical one and suffers from similar drawbacks. It provides no indication what order of components could work, and correctness proofs again boil down to classical ones. 

Attempts with a more complexity-theoretic angle consider restricted frameworks that capture interesting classes of instantiations, establish lower bounds for algorithms within the framework that solve certain computational problems, and show connections between approximability of optimization problems and their expressibility within the framework~\cite{KarpH67,Edmonds71,KorteL81,HelmanR85,Helman89,Woeginger00,BorodinNR03,AngelopoulosB04,DavisI09,AlekhnovichBBIMP11,Buresh-OppenheimDI11,JuknaS20}. 

A potential research direction is to formally capture the approach here as well as greedy-stays-ahead arguments, and to investigate their relationship in general. See also Section~\ref{sec:monotonicity}.

\paragraph{Organization.}
The next three sections develop the three instantiations mentioned above. Each time we review the standard dynamic program for the general problem and derive from it the well-known greedy algorithm for the restricted problem. Readers who want to quickly get to the core, can skip the reviews and jump to Section~\ref{sec:is:greed}, and then to Sections~\ref{sec:ks:greed} and \ref{sec:sssp:greed}. Section~\ref{sec:monotonicity} takes a closer look at the various monotonicity requirements involved, which relates to the open question.

Sections 2 to 4 each include the formal specifications of the optimization problems involved -- the general problem as well as the restricted one. All three problems have value versions (which output the value of an optimal solution) and argument versions (which output a valid setting of $y_1, \dots, y_n$ that yields the optimal value under $f_x$ for a given instance $x$). As the derivations of the greedy algorithms for the argument versions only rely on the dynamic programs for the value versions, we specify the value versions for the general problems and the argument versions for the restricted problems.

\section{Interval Scheduling}

Also known as activity selection or scheduling classes, \is\ aims to select from a given list of weighted intervals a non-overlapping subset of maximum total weight. For reasons of similarity with \ks, we will use the term ``value" rather than ``weight". Without loss of generality, the values are nonnegative. In regard to the alternate names, we refer to the start time and finish time of an interval. For ease of exposition, we consider non-empty half-open intervals. 

\newpage

\problemspec%
{\is\ (value version)}%
{}%
{nonempty interval $I_i = [s_i , f_i) \subseteq \RR$ with start time $s_i$ and finish time $f_i$ for $i \in [n]$ \\ value $v_i \in \RR_{\ge 0}$ for $i \in [n]$}%
{}%
{$\max\left\{\sum_{i \in S} v_i : S \subseteq [n] \wedge (\forall i \ne j \in S) \, I_i \cap I_j = \emptyset \right\}$}%
{}

The input can be viewed as consisting of the components $x_i=(s_i,t_i,v_i)$ for $i \in [n]$. We need to make a binary decision $y_i$ about each component $x_i$, where $y_i$ indicates whether interval $I_i$ is included. Alternately, the problem can be modeled as successive non-binary decisions as to which interval to include next (if any). We adopt the former view. 

\subsection{Dynamic program} 

We consider component $x_1$, determine the optimum under the two choices for $y_1$, and take the best of the two. For either choice of $y_1$, the optimum value can be expressed in terms of the solution of an instance given by a subsequence of $x_1,\dots,x_n$: the subsequence $x_2,\dots,x_n$ for $y_1=0$, and the subsequence of the indices $j$ with $I_1 \cap I_j = \emptyset$ for $y_1=1$. Repeated application of this reduction yields a recursive algorithm, which we can memoize. Since the order of the components within a sequence does not affect the outcome, the subinstances can all be described by subsets $J \subseteq [n]$. The crucial question is how many distinct subsets arise throughout the recursion.

For an arbitrary order of the components, the number can be $2^{\Omega(n)}$, as illustrated by the example in Figure~\ref{fig:is:bad-order}, in which $n=2m$ and the values remain unspecified.

\begin{figure}[h]
\centering 
\begin{tikzpicture}[scale=0.7]  
\draw[] (4,0) -- (15.45,0) ; 
\draw[dashed] (15.5, 0) -- (17.5, 0); 
\draw[] (17.5, 0) -- (21, 0);
\draw[-latex] (21,0) -- (22,0) node[above, near end] {$\RR$ \; }; 
\foreach \x in {1,2,3}{
  \draw[[-)] (4*\x,2) -- (4*\x+2,2) node[above,midway]{$I_{\x}$};
  \draw[[-)] (4*\x+1,1) -- (4*\x+3,1) node[above,midway]{$I_{m+\x}$};
}
\draw[] (16.5,2) -- (16.5,2) node {$\dots$};
\draw[[-)] (18,2) -- (20,2) node[above,midway]{$I_{m}$};
\draw[] (16.5,1) -- (16.5,1) node {$\dots$};
\draw[[-)] (19,1) -- (21,1) node[above,midway]{$I_{2m}$};
\end{tikzpicture}
\caption{Ordering for \is\ instance yielding $2^{\Omega(n)}$ distinct subinstances.}\label{fig:is:bad-order}
\end{figure}

At level $m$ of the recursion tree for the instance in Figure~\ref{fig:is:bad-order}, every possible subset $J$ of $\{m+1,\dots,2m\}$ arises as the remaining subinstance for a valid choice of decisions $y_1, \dots, y_m$, namely the choice $y_i = 0$ if $m+i \not\in J$, and $y_i=1$ otherwise. Thus, the number of distinct subinstances is at least $2^m = 2^{n/2}$. 

However, when we order the components earliest start time first, every subset $J$ in the recursion tree has the structure of a suffix of $[n]$, of which there are only $n+1$. This is because when we apply the reduction to a subinstance corresponding to the suffix $J \doteq \{i, \dots, n\}$ (initially $i=1$), either choice of $y_i$ merely removes a prefix from the suffix $J$: the prefix of length 1 for $y_i=0$, and the prefix consisting of all intervals with start time before $f_i$ for $y_i=1$. This leads to a 1-dimensional table $\OPT$ of subinstance solutions: For $i \in [n+1]$
\begin{equation}\label{eq:is:OPT}
\OPT(i) \doteq \text{optimal value for instance } x_i, \dots, x_n.
\end{equation}
The reduction yields the following Bellman equations:
\begin{equation}\label{eq:is:bellman}
\OPT(i) = \left\{ \begin{array}{ll}
    \max(\OPT(i+1),v_i+\OPT(\mathrm{next}(i)) & \text{ for } i \in [n] \\
    0 & \text{ for } i=n+1,
    \end{array} \right.
\end{equation}
where $\mathrm{next}(i)$ denotes the index of the first component $j$ whose start time $s_j$ comes at or after the finish time $f_i$ of component $i$, or $n+1$ if no such component exists. In symbols, 
\begin{align}\label{eq:is:next}
\mathrm{next}(i) \doteq \min\left\{ k \in \{i,\dots,n+1\}: s_k \ge f_i\right\},
\end{align}
where $s_{n+1} \doteq \infty$ is used as a sentinel. Note that $\mathrm{next}(i)$ can be computed in time $\Theta(\log n)$ using binary search when the components are ordered by start time. 

Sorting the components and then evaluating \eqref{eq:is:bellman} in reverse order yields the correct answer for the value version, $\OPT(1)$, in time $\Theta(n \log n)$. For future reference, here is pseudocode (assuming access to $\mathrm{next}$).

\begin{algorithm}[H]
\caption{Pseudocode for \is\ (value version)}\label{alg:is}
\begin{algorithmic}
   \State Sort components such that $s_i \le s_{i+1}$ for each $i \in [n-1]$
   \State $\OPT(n+1) \gets 0$
   \For{$i=n$ down to $1$} 
     \State $\OPT(i) \gets \max(\OPT(i+1),v_i+\OPT(\mathrm{next}(i)))$
  \EndFor
  \State \textbf{return} $\OPT(1)$
\end{algorithmic}
\end{algorithm}

\vspace{-1em}

Once we have computed the $\OPT$ table, a solution to the argument version can be retrieved in a second pass over the table, using the Bellman equations to determine which decisions are optimal. The second pass proceeds in the order of the components and 
takes $\Theta(n)$ time, resulting in a total running time of $\Theta(n \log n)$ for the argument version.  For future reference, we again include pseudocode (assuming access to $\mathrm{next}$ and $\OPT$).

\begin{algorithm}[H]
\caption{Pseudocode for \is\ solution retrieval}\label{alg:is:retrieval}
\begin{algorithmic}
    \State $S \gets \emptyset$; $i \gets 1$
    \While{$i \le n$} 
      \If{$\OPT(i) = \OPT(i+1)$} 
         \State $i \gets i+1$
      \Else
        \State $S \gets S \cup \{i\}$; $i \gets \text{next}(i)$
      \EndIf
   \EndWhile
   \State \textbf{return} $S$
\end{algorithmic}
\end{algorithm}

\subsection{Greedy algorithm}
\label{sec:is:greed}

We now restrict to the special cases where all values are 1. 

\problemspec%
{\uis\ (argument version)}%
{}%
{nonempty interval $I_i = [s_i , f_i) \subseteq \RR$ with start time $s_i$ and finish time $f_i$ for $i \in [n]$}%
{}%
{$S \subseteq [n]$ such that $|S|$ is maximized and $(\forall i \ne j \in S) \, I_i \cap I_j = \emptyset$}%
{}

We derive a greedy algorithm for the unit-value setting from the above dynamic program for the general setting. Consider the components ordered earliest start time first, as in the dynamic program. The recursive cases of the Bellman equations \eqref{eq:is:bellman} can be simplified using the precondition $v \equiv 1$ and  rewritten as
\begin{equation}\label{eq:is:bellman:rewritten}
\OPT(i) = \max(1+\OPT(\mathrm{next}(i)),\OPT(i+1)).
\end{equation}
Repeatedly applying \eqref{eq:is:bellman:rewritten} to expand the last $\OPT$ term on the right-hand side, and ultimately using the base case $\OPT(n+1)=0$ yields

\begin{eqnarray}
\OPT(i) & = & \max\Big( \, 1+\OPT(\mathrm{next}(i)), \nonumber \\
 & & \;\;\;\;\;\;\;\;\;\,\, 1+\OPT(\mathrm{next}(i+1)), \nonumber \\
 & & \;\;\;\;\;\;\;\;\;\,\,1+\OPT(\mathrm{next}(i+2)), \label{eq:is:expanded} \\ 
 & & \;\;\;\;\;\;\;\;\;\;\;\;\;\;\;\;\;\;\;\;\;\;\;\;\;\;\;\;\;\vdots \nonumber \\
 & & \;\;\;\;\;\;\;\;\;\,\,1+\OPT(\mathrm{next}(n)) \,\, \Big). \nonumber
\end{eqnarray} 
Each term $1+ \OPT(\mathrm{next}(j))$ in \eqref{eq:is:expanded} corresponds to the situation where interval $I_j$ is the first interval in the given order that is selected from among $I_i ,\dots, I_n$.
We can rewrite \eqref{eq:is:expanded} as
\begin{eqnarray}
 \OPT(i)
 & = & 1 + \max_{i \le j \le n}(\OPT(\mathrm{next}(j))) \label{eq:is:special} \\
 & = & 1 + \OPT(\min_{i \le j \le n}(\mathrm{next}(j))) \label{eq:is:monotone} \\
 & = & 1 + \OPT(\mathrm{next}(i^*)), \label{eq:is:arg}
 \end{eqnarray}
where 
\begin{equation}\label{eq:is:star}
i^* \doteq \arg \min_{i \le j \le n} \mathrm{next}(j).
\end{equation}
Step~\eqref{eq:is:special} follows by extracting the constant 1 and is the one and only step that hinges on the restriction of the special case. Step~\eqref{eq:is:monotone} holds because $\OPT$ is non-increasing, a property that follows immediately from its defining equation \eqref{eq:is:OPT}: Increasing the argument of $\OPT$ means shrinking the set of components that can be used for the underlying maximization problem. Step~\eqref{eq:is:arg} merely makes the argument that achieves the maximum explicit, which is what definition~\eqref{eq:is:star} of $i^*$ does. (It is irrelevant how ties for $i^*$ are broken; for concreteness, we pick the smallest among the tied candidates.)

By definition \eqref{eq:is:next}, the value of $\mathrm{next}(j)$ is determined by the finish time $f_j$ of $I_j$, and is non-decreasing as a function of $f_j$ since increasing $f_j$ shrinks the set of components $k$ of which we take the minimum. Thus, $i^*$ is the (first) index of an interval $I_j$ with the earliest \emph{finish} time among $I_i, \dots, I_n$. The interpretation of the terms in \eqref{eq:is:expanded} then tells us that an optimal solution is to select interval $I_{i^*}$ and then continue with the remaining intervals that do not overlap with $I_{i^*}$. That is to say, we greedily consider the intervals earliest finish time first, which is exactly what the classical greedy algorithm does. 

For completeness, we include pseudocode for an efficient implementation. 
\begin{algorithm}[H]
\caption{Pseudocode for \uis\ (argument version)}\label{alg:uis}
 \begin{algorithmic}
    \State Sort components such that $f_i \le f_{i+1}$ for each $i \in [n-1]$
    \State $S \gets \emptyset$; $f \gets -\infty$
    \For{$i=1$ to $n$} 
      \If{$s_i \ge f$} 
         \State $S \gets S \cup \{i\}$; $f \gets f_i$
      \EndIf
   \EndFor
   \State \textbf{return} $S$
 \end{algorithmic}
\end{algorithm}

\vspace{-1em}

After the initial sorting, which takes time $\Theta(n \log n)$, the remainder of the algorithm runs in time $\Theta(n)$, resulting in an overall running time of $\Theta(n \log n)$.

\subsection{Order considerations}

We end our discussion of \is\ with a few additional remarks about the different orders involved. The order of components that we mention is the one for applying the reduction that underlies both paradigms. In the dynamic program, this order is used in the memoized process and in the second phase (solution retrieval) of the iterative implementation (see Algorithm~\ref{alg:is:retrieval}). The first phase of the iterative implementation considers the components in the reverse order (see Algorithm~\ref{alg:is}).

The reverse order in the dynamic program for \is\ is earliest start time last. Upon reversal of the time line, this becomes earliest finish time first. The orientation of the time line does not affect correctness. As an increasing loop variable is more natural than a decreasing one, rather than sorting components earliest deadline first as in Algorithm~\ref{alg:is}, typical implementations of the first phase of the dynamic program for \is\ sort the components earliest finish time first. This agrees with the greedy order for \uis, but the agreement is a coincidence. When teaching greed before dynamic programming, picking the greedy order for \uis\ as the order in the first iterative phase of the dynamic program for \is\ requires justification. The justification for the greedy order is correctness, whereas the justification for the order in the dynamic program is efficiency. 

To complete the picture, we analyze ordering the components earliest finish time first in the dynamic program for \is. The subsets $J$ are no longer merely suffixes of $[n]$, but still have enough structure to keep the number of subinstances small, albeit larger than for earliest start time first. Each $J$ now has the form $\{ j \in \{i, \dots, n\} : s_j \ge f_\ell\}$ for $0 \le \ell < i \le n$, where $f_0 \doteq -\infty$ acts as a sentinel. The parameter space is therefore 2-dimensional and of size $O(n^2)$. Figure~\ref{fig:is:earliest-start} describes instances for which $\Omega(n^2)$ of the combinations $(\ell,i)$ actually occur. This means that the number of subinstances for earliest finish time first is $\Theta(n^2)$ as opposed to $\Theta(n)$ for earliest start time first, resulting in a running time of $\Theta(n^2)$ in lieu of $\Theta(n \log n)$. 

\begin{figure}[h]
\centering 
\begin{tikzpicture}[scale=0.7]  
\draw[] (2,0) -- (11.45,0) ; 
\draw[dashed] (11.5, 0) -- (14, 0); 
\draw[] (14, 0) -- (20, 0); 
\draw[dashed] (20, 0) -- (22, 0); 
\draw[-latex] (22,0) -- (24,0) node[above, near end] {$\RR$ \; };      
\foreach \x in {1,2,3}{
    \draw[[-)] (3*\x,6) -- (3*\x+2,6) node[above,midway]{$I_{\x}$};
    \draw[[-] (3*\x+1,6-\x) -- (11.45,6-\x) node[above,midway]{$I_{m+\x}$};
    \draw[dashed] (11.5,6-\x) -- (14,6-\x);
    \draw[-)] (14,6-\x) -- (16.5+\x,6-\x);
}
\draw[] (12.75,6) -- (12.75,6) node {$\dots$};
\draw[[-)] (14.5,6) -- (16.5,6) node[above,midway]{$I_{m}$};
\draw[] (16.5,2.2) -- (16.5,2.2) node {$\vdots$};
\draw[[-] (15.5,1) -- (20,1) node[above,midway]{$I_{2m}$};
\draw[dashed] (20, 1) -- (22, 1);
\draw[-)] (22,1) -- (22.5,1);
\end{tikzpicture}
\caption{\is\ instance where earliest start time first yields $\Theta(n^2)$ distinct subinstances.}\label{fig:is:earliest-start}
\end{figure}

The instances in Figure~\ref{fig:is:earliest-start} have even $n=2m$, intervals ordered earliest finish time first, and unspecified values. For each $i \in [m]$, all subinstances with parameters $(\ell,i)$ for $0 \le \ell \le i-1$ are present at level $i-1$ of the recursion tree. They correspond to subsets of the form $J = \{i, \dots, m\} \cup \{m+\ell+1,\dots,2m\}$ and occur when we set $y_{\ell}=1$ (for $\ell \ge 1$) and $y_{\ell+1} = \dots = y_{i-1} = 0$.  There are $\binom{m+1}{2} = \Theta(n^2)$ such subinstances in total.

\section{Knapsack}

Given a list of items characterized by their weights and values, we want to select a subset of maximum total value that does not exceed the weight limit of our knapsack. Weights are intended to be nonnegative; values are nonnegative without loss of generality. In order to make the problem amenable to dynamic programming, we need to constrain either the weights or the values to be discrete. We opt for the weights.

\problemspec%
{\ks\ (value version)}%
{}%
{item $i$ with weight $w_i \in \ZZ_{\ge 0}$ and value $v_i \in \RR_{\ge 0}$ for $i \in [n]$ \\ weight limit $W \in \ZZ_{\ge 0}$}%
{}%
{$\max\left\{\sum_{i \in S} v_i : S \subseteq [n] \wedge \sum_{i \in S} w_i \le W \right\}$}%
{}

Similar to \is, the input can be viewed as consisting of the components $x_i=(w_i,v_i)$ for $i \in [n]$ and the additional quantity $W$. We need to make a binary decision $y_i$ about each component $x_i$, where $y_i$ indicates whether item $i$ is included in the knapsack. Alternately, the sequence of binary decisions can be replaced by successive non-binary decisions as to which item to include next (if any). We again adopt the former view. 

\subsection{Dynamic program} 

Given a decision for the first item, it remains to solve the instance consisting of items 2 through $n$ with weight limit $W$ in case we do not include item 1, and the same instance but with a reduced weight limit of $W - w_1$ in case we include item 1 (which is only an option if $w_1 \le W$). The subinstances that arise through repeated application of the reduction consist of a suffix of the given list of items, and a nonnegative integer at most $W$ as the weight limit. The solutions can be represented in a matrix $\OPT$. For $i \in [n+1]$ and $w \in \{0,\dots,W\}$
\begin{equation}\label{eq:ks:OPT}
\OPT(i,w) \doteq \text{optimal value for components } x_i, \dots, x_n \text{ and weight limit } w.
\end{equation}
The Bellman equations are
\begin{equation}\label{eq:ks:bellman}
\OPT(i,w) = \left\{ \begin{array}{ll}
    \max\left(\OPT(i+1,w), \underbrace{v_i+\OPT(i+1,w-w_i)}_{\text{considered only if } w_i \le w}\right) & \text{ for } i \in [n] \\
    0 & \text{ for } i=n+1.
    \end{array} \right.
\end{equation}
We evaluate \eqref{eq:ks:bellman} row-by-row from row $i=n$ to row $i=1$. This yields the optimal value, $\OPT(1,W)$, in time $\Theta(n \cdot W)$. A subset $S \subseteq [n]$ realizing the optimal value can be retrieved from the $\OPT$ table in a similar fashion as in Algorithm~\ref{alg:is:retrieval}, based on \eqref{eq:ks:bellman} in lieu of \eqref{eq:is:bellman}. The retrieval takes an additional $\Theta(n)$ steps, resulting in an overall running time of $\Theta(n \cdot W)$ for the argument version of $\ks$.

\subsection{Greedy algorithm}
\label{sec:ks:greed}

We now restrict to the special cases where all values are 1. 

\problemspec%
{\uks\ (argument version)}%
{}%
{item $i$ with weight $w_i \in \ZZ_{\ge 0}$ for $i \in [n]$ \\ weight limit $W \in \ZZ_{\ge 0}$}%
{}%
{$S \subseteq [n]$ such that $|S|$ is maximized and $\sum_{i \in S} w_i \le W$}%
{}

We derive a greedy algorithm for the unit-value setting from the above dynamic program for the general setting. The recursive cases of the Bellman equations \eqref{eq:ks:bellman} can be simplified using the precondition $v \equiv 1$ and rewritten as
\begin{equation}\label{eq:ks:bellman:rewritten}
\OPT(i,w) = \max\left(\underbrace{1+\OPT(i,w-w_i)}_{\text{considered only if }w_i \le w},\OPT(i+1)\right).
\end{equation}
Repeatedly applying \eqref{eq:ks:bellman:rewritten} to expand the last $\OPT$ term on the right-hand side and ultimately the base case $\OPT(n+1,\cdot)=0$ yields
\begin{equation}\label{eq:ks:expanded}
\OPT(i,w) = \max\left\{ 1+\OPT(j+1,w-w_j) : i \le j \le n \wedge w_j \le w \right\},
\end{equation}
where the term for $j$ corresponds to the choice of $j$ as the first item from among $i,\dots,n$ to be included. The maximum of the empty set is considered zero, corresponding to the case where none of the items $i$ through $n$ has weight at most $w$. 

By its definition \eqref{eq:ks:OPT}, $\OPT(i,w)$ is non-increasing in $i$ and non-decreasing in $w$. In general, these monotonicity properties in the individual arguments are insufficient to determine the optimal choice for $j$ in \eqref{eq:ks:expanded}. However, if $w_i$ happens to be the minimum of $w_i, \dots, w_n$, then 
\begin{equation*}
\OPT(i,w-w_i) \ge \OPT(i,w-w_j) \ge \OPT(j,w-w_j)
\end{equation*}
holds for any $j \in \{i,\dots,n\}$, where the first inequality follows from the monotonicity of $\OPT$ with respect to its second argument, and the second inequality from the monotonicity of $\OPT$ with respect to its first argument. Thus, if we order the items lightest weight first (with ties broken arbitrarily), $j=i$ always is an optimal choice in \eqref{eq:ks:expanded} (provided any valid choice exists). Here is pseudocode for a straightforward $\Theta(n \log n)$ implementation of the resulting greedy strategy, using $w_{n+1} \doteq \infty$ as a sentinel. 
\begin{algorithm}[H]
\caption{Pseudocode for \uks\ (argument version)}\label{alg:uks}
 \begin{algorithmic}
    \State Sort items such that $w_i \le w_{i+1}$ for each $i \in [n-1]$
    \State $S \gets \emptyset$; $w \gets W$; $i \gets 1$
    \While{$w_i \le w$} 
        \State $S \gets S \cup \{i\}$; $w \gets w-w_i$
   \EndWhile
   \State \textbf{return} $S$
 \end{algorithmic}
\end{algorithm}

\vspace{-1em}

Due to the sorting, Algorithm~\ref{alg:uks} takes time $\Theta(n \log n)$. For completeness, we point out that the running time can be improved to $\Theta(n)$. The output $S^*$ is the longest prefix of the sorted order that has total weight at most $W$. The length of the prefix $S^*$ can be found in linear time using binary search and linear-time selection.

\section{Shortest Paths}

Given a digraph $G$, edge lengths, and two vertices, $s$ and $t$, we want to find a shortest path from $s$ to $t$ in $G$, or report that no such path exists. Nonexistence happens if $G$ has no path from $s$ to $t$ at all. When negative edge lengths are allowed, nonexistence can also occur because there is a path from $s$ to $t$ but every such path can be made shorter. We distinguish between the two cases of nonexistence in the notion of distance.

\begin{definition}[distance] Let $G=(V,E)$ be a digraph with edge lengths $\ell: E \to \RR$, and $s,t \in V$. The distance from $s$ to $t$ is
\[ d(s,t) \doteq \left\{ \begin{array}{ll}
\infty & \text{ if there is no path from $s$ to $t$} \\
-\infty & \text{ if there is a path from $s$ to $t$ but no shortest one} \\
\min\{ \ell(P) : P \text{ path } s \rightsquigarrow t \} & \text{ otherwise},
\end{array} \right.
\]
where $\ell(P) \doteq \sum_{e \in P} \ell(e)$ denotes the length of the path $P$ (viewed as a sequence of edges) and $s \rightsquigarrow t$ denotes that the path starts at $s$ and ends at $t$.
\end{definition}

The value version of the problem amounts to computing the distance $d(s,t)$. Typical algorithms implicitly compute the distance for more pairs of vertices than just $(s,t)$. We focus on the single-source variant of the problem, which explicitly outputs the distance for a given source $s$ and all possible targets $t$.

\problemspec%
{\sssp\ (value version)}%
{}%
{digraph $G=(V,E)$, $\ell: E \to \RR$, $s \in V$}%
{}%
{$d(s,t)$ for every $t \in V$}%
{}

In the argument version, we need to additionally output a shortest path from $s$ to $t$ whenever one exists. We impose two natural constraints. First, we want each output path to be \emph{simple}, i.e., to have no repeated vertices. This is possible because any path $P$ from $s$ to $t$ that repeats a vertex contains a cycle as a subpath. Since $d(s,t) \ne -\infty$, the cycle has to have nonnegative length and can be cut out from $P$ without increasing the path length. Starting with an arbitrary shortest path from $s$ to $t$, a finite number of such cuts yields a simple shortest path, as desired. Second, the principle of optimality implies that any prefix of a shortest path from $s$ to $t$ is a  shortest path from $s$ to some (other) vertex. We impose such relationships on the single-source output. More precisely, whenever $P$ is the path output for $t$ then, for any vertex $v$ on the path, the subpath of $P$ from $s$ to $v$ is the path output for $v$. 

Combined, the two constraints are equivalent to the shortest paths forming a tree with root $s$ and edges oriented away from the root. Such a collection of paths can be described by a predecessor function $p: F \setminus \{s\} \to F$, where $F$ denotes the set of vertices $t \in V$ with finite $d(s,t)$, and $p(t)$ symbols the predecessor of $t$ on the chosen shortest path from $s$ to $t$. A simple shortest path from $s$ to $t$ can be retrieved in reverse order by starting from $t$ and repeatedly applying $p$ until we reach $s$.

\subsection{Dynamic program}
\label{sec:sssp:dp}

Consider a nonempty path $P$ from $s$ to $t$. We apply the principle of optimality to the subpath of $P$ without its last edge, and try all possibilities for the last edge. The approach yields the following Bellman equations for the distances $d(s,t)$ for every vertex $t \ne s$ that is reachable from $s$:
\begin{equation}\label{eq:sssp:bellman:d}
d(s,t) = \min \{d(s,v) + \ell(v,t) : (v,t) \in E\}.
\end{equation}
These equations are all we need to know about the dynamic program to derive a greedy algorithm under the precondition of nonnegative edge lengths. The dynamic program, known as Bellman-Ford, computes the distances and solves the argument version in time $\Theta(n(n+m))$, where $n \doteq |V|$ and $m \doteq |E|$.

\subsection{Greedy algorithm}
\label{sec:sssp:greed}

We now restrict attention to digraphs with nonnegative edge lengths. In this setting, distances cannot be negative. We again consider the single-source variant. 

\problemspec%
{\nsssp\ (argument version)}%
{}%
{digraph $G=(V,E)$, $\ell: E \to \RR_{\ge 0}$, $s \in V$}%
{}%
{$d(s,t)$ for every $t \in V$ \\
predecessor $p(v)$ of $v$ on a simple shortest path $s \rightsquigarrow t$ for every $s \ne t \in V$ with finite $d(s,t)$}%
{}

We derive Dijkstra's greedy algorithm from the Bellman equations \eqref{eq:sssp:bellman:d} for the distances. We need to determine the order to compute the distances $d(s,t)$ for all reachable $t \in V$. We start with $t=s$. Since edge lengths are nonnegative, the empty path is optimal, witnessing that $d(s,s)=0$ and leaving $p(s)$ undefined.

In every subsequent step, let $T$ denote the set of vertices $t$ for which we have already determined $d(s,t)$. If there are no reachable vertices outside of $T$, we know that all remaining distances are $\infty$ and can stop. Otherwise, consider any reachable vertex $t \in \overline{T} \doteq V \setminus T$. For analysis purposes, start from the single term $d(s,t)$ at level 0. Obtain each next level from the previous one by rewriting every term of the form $d(s,v)$ for $v \in \overline{T}$  as
\begin{equation*}
d(s,v) \to \min \{d(s,u) + \ell(u,v) : (u,v) \in E\}
\end{equation*}
based on \eqref{eq:sssp:bellman:d}.
At level $n$, the resulting expression is the minimum of the following terms:
\begin{itemize}
\item[(a)] $d(s,u) + \ell(u,t)$ for all $u \in T$ with $(u,t) \in E$. These terms correspond to expansions that reach a vertex $u$ inside $T$ at the first level.
\item[(b)] $d(s,u) + \ell(u,v)$ + $\ell(P)$ for some $(u,v) \in E$ with $u \in T$, $v \in \overline{T}$, and some nontrivial path $P: v \rightsquigarrow t$. These terms correspond to expansions that reach a vertex $u$ inside $T$ but not at the first level.
\item[(c)] $d(s,u) + \ell(P)$ for some $u \in \overline{T}$ and some path $P$ inside $\overline{T}$ containing $n-1$ edges. These terms correspond to expansions where no level reaches a vertex $u$ inside $T$. 
\end{itemize} 
The minimum over \emph{all} reachable $t \in \overline{T}$ is achieved by a term of type (a). This follows from the non-negativity of the edge lengths:
\begin{itemize}
\item The term $d(s,u) + \ell(u,v)$ + $\ell(P)$ of type (b) is at least $d(s,u) + \ell(u,v)$, which is a term of type (a) as $v \in \overline{T}$ is a possible choice for $t$ in (a).
\item The term $d(s,u) + \ell(P)$ of type (c) can be ignored because it corresponds to paths with $P$ as a subpath, and $P$ contains a repeated vertex since it has $n-1$ edges and lies entirely in the set $\overline{T} \subseteq V \setminus \{s\}$ of size at most $n-1$.
\end{itemize} 
The above analysis shows that for 
\begin{equation}\label{eq:sssp:greedy}
(u^*,v^*) \doteq \arg \min_{(u,v) \in E \cap T \times \overline{T}}(d(s,u) + \ell(u,v)),
\end{equation}
we can compute $d(s,v^*)$ from the distances we already know, namely, as $d(s,v^*) = d(s,u^*) + \ell(u^*,v^*)$. Thus, we pick $t=v^*$ as the next vertex in the order, include $t$ into $T$, and set $p(v^*) = u^*$.

To efficiently calculate \eqref{eq:sssp:greedy}, we maintain the vertices $v \in \overline{T}$ in a priority queue, keyed on $\lambda(v) \doteq \min( \{\infty\} \cup \{d(s,u) + \ell(u,v) : (u,v) \in E\} )$. The iteration for $t$ requires one minimum extraction and $\deg^{+}(t)$ key updates, where $\deg^{+}(t)$ denotes the out-degree of $t$. In total, we need $n \doteq |V|$ minimum extractions and $\sum_{t \in V} \deg^{+}(t) = |E| \doteq m$ key updates. An implementation based on binary heaps runs in time $\Theta((n+m) \log n)$ and an implementation based on Fibonacci heaps in time $\Theta(m + n \log n)$.

\section{Monotonicity Properties}
\label{sec:monotonicity}

We end with some reflections on the monotonicity conditions we used and compare them with monotonicity requirements in other works, in particular the formulations based on the categorical calculus of relations.

The principle of optimality~\cite{Bellman54} that underlies dynamic programming can be viewed as a monotonicity property: Better solutions to subinstances of a given input cannot worsen the solution to the given input. The framework of \cite{BirdM93} contains an explicit requirement that the combining function of the subinstances be monotone with respect to the objective function. Under this condition, the framework produces a correct dynamic program. 

To obtain a correct greedy algorithm, the framework needs an additional requirement, which can also be viewed as a monotonicity condition in the categorical calculus of relations: 
\begin{equation}\label{eq:rel}
S \cdot \beta \subseteq \beta \cdot R,
\end{equation}
where $\cdot$ denotes composition of relations, $S$ captures the local criterion, $R$ the global objective, and $\beta$ is the prefix relation between partial and full solutions. The inclusion \eqref{eq:rel} states that if $p_1$ and $p_2$ are partial solutions that agree on all but the last step, and the local criterion prefers $p_1$ over $p_2$, then for any full solution $y_2$ that extends $p_2$, there exists a full solution $y_1$ extending $p_1$ such that $y_1$ is at least as good as $y_2$ with respect to the global objective. \cite{Curtis03} refers to requirement \eqref{eq:rel} as  ``better global", and develops similar results under the weaker ``best global" requirement, which only imposes the above condition for prefixes $p_1$ in which the last step is \emph{optimal} with respect to the local criterion. In fact, an even weaker requirement suffices, namely, imposing the condition on prefixes $p_1$ that are obtained by successive locally-optimal choices. This last requirement merely states that the local criterion guarantees optimality. 

\cite{Curtis03} also defines local variants of ``best" and ``better", where the global objective is replaced by a quality measure for partial solutions. They can be viewed as formalizations of greedy-stays-ahead in the general categorical framework of relational calculus. All four variants enable synthesizing a greedy algorithm within the framework, but need a local criterion $S$ that satisfies the requirement \eqref{eq:rel} or its alternates. The construction of $S$ and the proof that it works are left to the algorithm designer, as is how to (re)formulate a given computational problem in a way that $S$ exists.

Our monotonicity properties differ from the ones in the category theoretic setting. In the context of \is\ and \ks, we need to relate the optimal objective values for \emph{different} subinstances, e.g., for an instance and the instance with one component less. In contrast, the requirements in the category theoretic setting relate the objective values of different solutions to the \emph{same} subinstance.

For completeness, we mention that monotonicity properties have been used in other contexts to improve dynamic programs, without turning them into greedy algorithms. An example is the monotonicity of the break points in the classical dynamic program for optimal binary search trees, which allows shaving off a linear factor in the running time and relates to totally monotone matrices \cite{Knuth71,Yao82,AggarwalKMSW87,BeinGLZ09}. Such monotonicity properties differ from ours as they refer to arguments whereas ours refer to values of the objective function. 

\section*{Acknowledgements}

I would like to thank Allan Borodin, Sanjoy Dasgupta, Jon Kleinberg, Tim Roughgarden, and Eva Tardos for feedback and encouragement. I am also grateful to (former) students Tom Watson, Drew Morgan, Nicollas Sdroievski, and Ahmed Shaaban for relevant discussions.

\bibliographystyle{alpha}
\bibliography{references}

\end{document}